# Towards Supporting Sustainable Grocery Shopping through Joyful Technology: Annotated Portfolio of Speculative Ideas

**Gözel Shakeri**
University of Oldenburg
Oldenburg, Germany
Gozel.Shakeri@uol.de

**Frederike Jung**
OFFIS
Oldenburg, Germany
Frederike.Jung@offis.de

**Ferran Altarriba Bertran**
Universitat de Girona
Girona, Spain
ferranaltarriba@gmail.com

**Daniel Fernández Galeote**
Tampere University
Tampere, Finland
daniel.fernandezgaleote@tuni.fi

**Adrian Friday**
Lancaster University
Lancaster, UK
a.friday@lancaster.ac.uk

## ABSTRACT

A third of greenhouse gas emissions are attributable to the food sector. A shift in dietary habits could reduce these by half. Engaging and empowering consumers is vital to this critical shift; yet, if we get the framing wrong, we might cause distress or eco-anxiety, impeding initial engagement as well as longer-term diet change. Evoking joy is a powerful yet under-explored motivator to overcome psychological barriers and support pro-environmental attitudes. This pictorial presents the outcomes of a one-day workshop as a series of speculative ideas in the form of an annotated portfolio, highlighting design qualities and interaction mechanisms that afford joy and sustainability in food choices. Our contribution will inspire HCI researchers and designers to reposition joy as a fundamental value to sustainability communication.

## CSS Concepts
• Human-centered computing → Information visualization; Interaction design; Visualization.

## MOTIVATION

A third of environmental degradation is attributable to the food sector [40]. Dietary change could reduce this by half [53]. Thus, eating more sustainably (i.e., consuming more plant-based, local, and seasonal foods) is of growing importance. Accordingly, nearly three-quarters of North-West Europeans [1, 12] think it is important for them to buy food that has a low environmental impact, yet only 7% regularly do [11, 55]. While there are clearly many system barriers beyond the individual consumer, engaging consumers in this change is more important than ever. Much research focuses on the provision of information when sustainable grocery shopping, in-store and online. Solutions range from sustainable recommender systems [57] and user-preference based systems [26] to incen-tivisation strategies [19, 37], and behavioural nudges [32, 37, 51]. However, there are two major issues with these approaches.

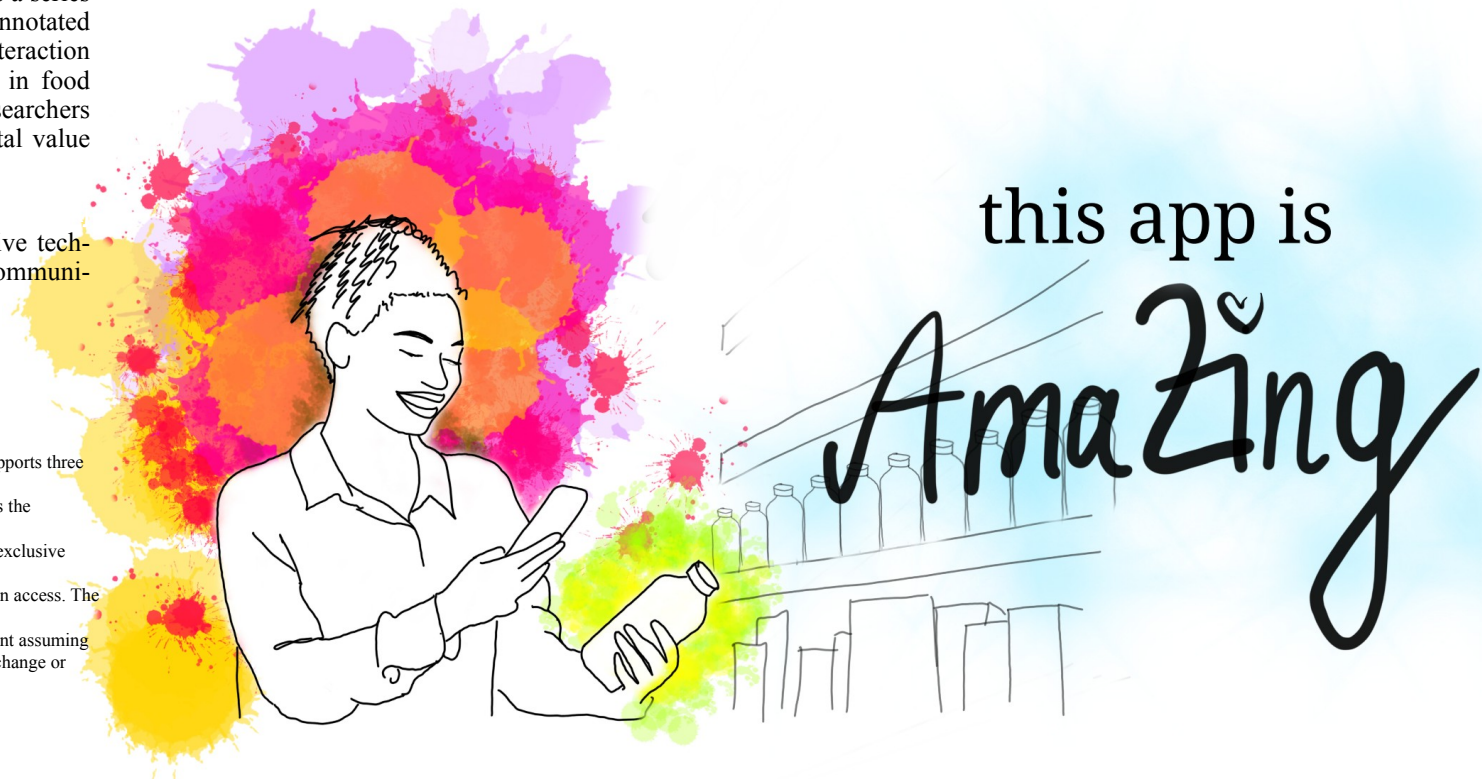

First, existing solutions side-line positive emotional involvement. A growing body of literature argues that the widely-held view of "lack of information" about climate change is not the main obstacle to engagement [31]; rather a lack of positive emotional involvement is. When exposed to environmental degradation [58], the pri-mary emotional reactions we may feel are a sense of diminished control and powerlessness [2,13,14]. Mechanisms aimed at relieving us from these negative feelings are denial, rational distancing, apathy, and delegation [25]. Thereby, emotions function as *antecedent* to engagement [47], obstructing sustainable behaviours. Additionally, emotions function as *consequence* of engagement. Individuals who are particularly invested in sustainability are further impacted by negative feelings such as anxiety, distress, dread, guilt, shame, and frustration [34, 59, 66], reinforced by the notion that engagement is intrinsically utilitarian and denies self-indulgence.

Ultimately, negative feelings prior and post pro-environmental behaviours, and the shift of responsibility onto the individual and away from system drivers including affordability and access causes disengagement with climate action altogether [41]. Interventions that only target factors such as knowledge, instead of attempting to overcome negative emotions are unlikely to cause significant changes in socially and culturally embedded high-impact eating habits [5, 39].

We posit that *joy*[1] may help consumers to find and maintain holistically meaningful ways to take action, thus promoting and sustaining pro-environmental efforts [25, 47, 56, 61]. Tools that elicit positive emotions, such as feeling amazed, cheerful, proud, and hopeful, have been shown to be significant predictors of purchase intentions, and can increase the probability of making a purchase [17, 48]. Overlooking the importance of joy when designing systems supporting sustainable food purchases may hinder their success [17, 47, 49].

Second, existing solutions utilise primarily (front-of-package) labels as a language to communicate a food's sustainability, focusing again on quantitative and symbolic representation rather than sparking joy. While eco-labels could elicit positive emotions such as pride over purchase behaviour [17, 36], eco-labels can lead to consumer confusion through information overload [63] as well as lack of information [20], decreased consumer trust and label credibility [21, 33], and ultimately in feelings of helplessness [34, 59, 66] once sustainable purchase intentions are abandoned [3, 16, 54]. More generally, the intention and impact of eco-labels and existing digital solutions do not extend beyond technology as guidance, as mediators, or as tools for reflection [29].

Despite potentially playing an important role in shaping people's climate conscious decision-making, joy is clearly not sufficient nor is there a one-size-fits-all approach [47]. The expression of joy depends on situational circumstances and individual variables, including one's personal understanding of joy [47]. Therefore, in this work, we explore what might be 'joyful elements' for sustainability signalling by a research-through-design workshop.

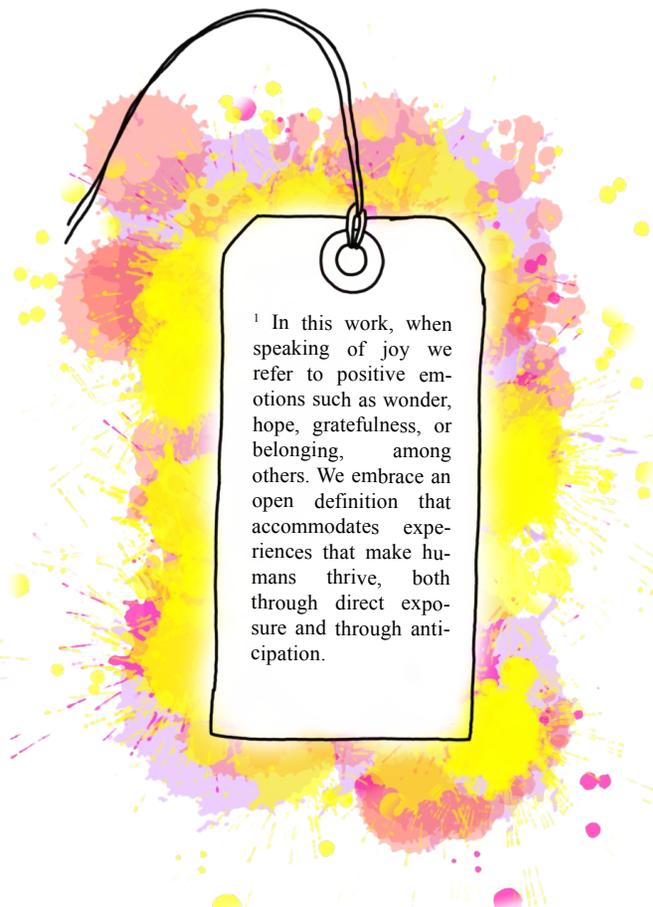

[1] In this work, when speaking of joy we refer to positive emotions such as wonder, hope, gratefulness, or belonging, among others. We embrace an open definition that accommodates experiences that make humans thrive, both through direct exposure and through anticipation.

The ubiquity of digital tools can cater to a multitude of factors by providing versatile media (e.g. apps, web-browser add-ons, Augmented Reality (AR) glasses) in versatile ways (e.g. in-store, online). Acknowledging the possibly pivotal role of digital technologies, we invited HCI researchers and designers to a workshop where we co-designed technology-mediated eco-labels which support consumers in pro-environmental decision-making. The overarching theme of the workshop was to rethink sustainability communication and to incorporate joy into it.

This paper presents the outcomes of conversations that took place during the workshop, aimed at imagining the future of digital sustainability support. In particular, our contribution is twofold: first, we contribute an annotated portfolio of five speculative ideas for eco-labels that utilise emerging technologies in ways that may be both *effective* and *joyful*. Second, we provide a close reading of these designs [9] to reflect on common qualities and themes to sustainability signalling. Aiming to 'reposition joy' as a fundamental aspect of sustainability signalling, this work presents researchers and designers in this field with tangible ideas on how to incorporate joy in sustainable grocery shopping.

**BACKGROUND**

Joy is key in pro-environmental behaviour and must be considered for productive engagement with climate change. For example, the way we communicate about animal-derived products often highlights positive emotions, triggering re-experiences of eating and enjoying foods, often in social circumstances [39] (e.g., family barbecue), thereby guiding consumption processes such as habits and perceived pleasure [15, 42, 50]. In contrast, communication about sustainable foods often lacks enjoyment, focusing instead on informational and utilitarian factors such as the sustainability of the ingredients or the greater good for society and the planet [39]. Consequently, sustainability becomes a moral obligation towards others rather than an intrinsic value. This, once framed as an extrinsic source of motivation, can tie the behaviour to contingent and external dynamics of reward and punishment [44].

Then, associated negative emotions may function as hindrance to future purchase behaviours. Evoking joy instead may be key to promoting productive engagement with climate change.

The single most used technology to support sustainable grocery shopping are food labels. The basic idea of food label technology is to provide information, enable comparison, and give feedback [10, 46], using either abstract or concrete information, which is presented visually or through text, giving complex information about a product's characteristics, such as animal welfare standards, environmental impacts, and ethical wage labour, in a simplified form to make it easier for consumers to make informed decisions [4, 8, 21, 28]. HCI examples of eco-labels on food are EcoPanel [65], Social Recipes [62], Envirofy [51], Nu-Food [38], Econundrum [46], and Food Qualculator [7]. These works investigated labels as a means of providing information tailored to users' own context and choices, without a direct focus on enabling joy.

Beyond being primarily informational, labels are traditionally static. They are either printed onto food packaging, visualised in sculptures [46], or displayed with a product during technology-mediated shopping [52]. Static solutions operate from a one-size-fits-all mentality; pro-environmental consciousness [25] however is a complex made up of environmental knowledge, values and attitudes, and emotional involvement. A single solution cannot cater to the multitude of know-ledges and aspects of joy. In HCI, attempts to create dynamic labels exist (e.g. [26, 51]), but they do not con-sider joy, ignoring a key aspect necessary for successful sustainability support when grocery shopping.

Finally, labels may elicit emotions, but often negative ones [17]. On the one hand, labels empower consumers to adhere to their sustainability goals, imbuing a sense of pride through accomplishment [17]. On the other hand, labels do not address the distress experienced in regards to climate change engagement, such as the sacrifice of personal pleasure over global interests, etc. Involving people's emotional responses more intentionally and actively may overcome these psychological (and societal) challenges.

In summary, there exists a gap in the literature which investigates *effective* and *joyful* sustainability support. Digital technological solutions, to a larger extent than printed packaging labels, have the ability to display a multitude of personalised or emotionally appealing content and allow the shopper to interact with sustainability information. HCI researchers play a crucial role in designing, creating, and evaluating effective visualisations that encourage sustainable food choices, positive grocery shopping experiences, and ultimately reduce greenhouse gas emissions. This pictorial presents a series of five speculative ideas in the form of an annotated portfolio, highlighting interesting aspects that afford joyful and sustainable food choices.

## WORKSHOP

This pictorial is based on the results of a workshop held at [omitted]. It was based on the concept of "choosing (with) joy" in which participants explored the idea of joy in eco-labels, sustainable food commu-nication, and technologies. After piloting the workshop with six participants in a prior setting, we then con-ducted the workshop in one day with 11 participants, including 5 of the co-authors. All participants worked in the field of HCI, as researchers or designers. Overall, the workshop consisted of three parts: Examining the state-of-the-art in sustainability communication; design-ing speculative, digital tools for sustainability communi-cation; and delving into the ideas and their underlying qualities.

Together with the participants we discussed surfaced overall themes to joyful, technology-mediated sustain-ability communication. After the workshop, the authors iterated this analysis. We revisited the common qualities and themes to sustainability signalling through a close reading of the speculative designs [9], and linked them back to existing literature, in an attempt to challenge and solidify the insights from the workshop. The result is a set of joyful speculative ideas, their design qualities and themes that inspire designers and HCI researchers to reclaim joy as a fundamental value for sustainable grocery shopping.

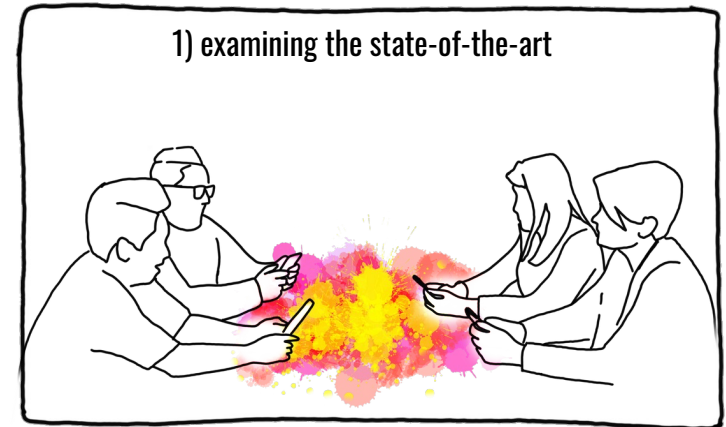

1) examining the state-of-the-art

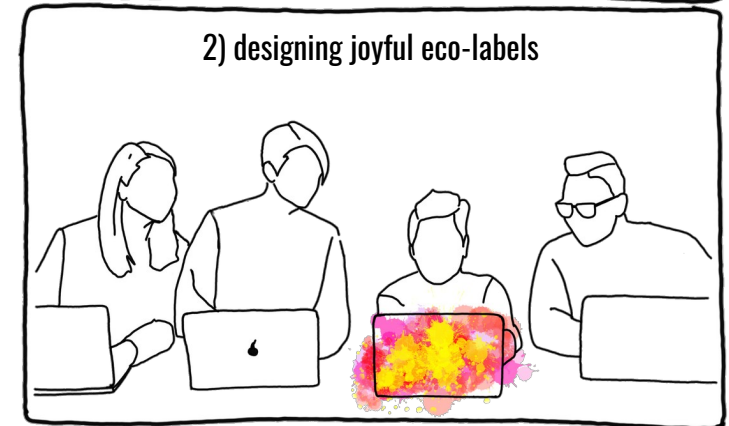

2) designing joyful eco-labels

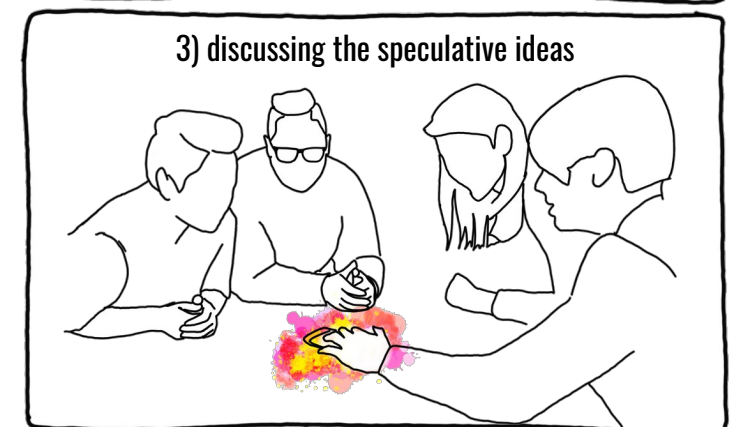

3) discussing the speculative ideas

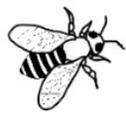
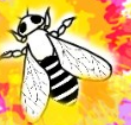
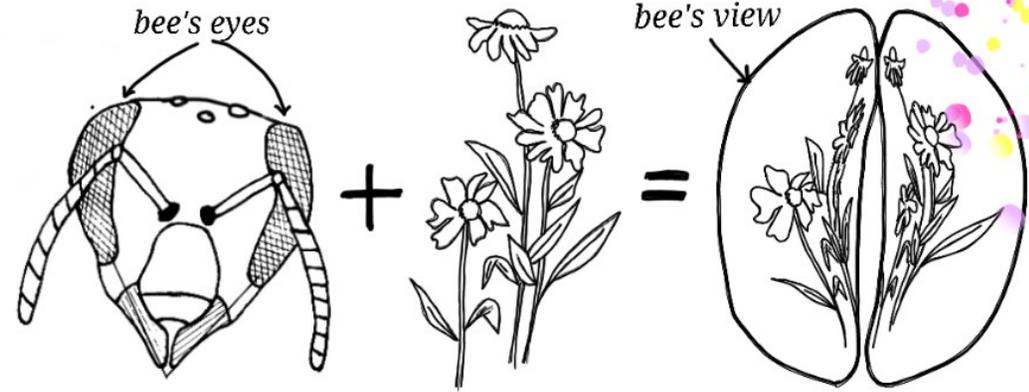

## BE(E)HIND THE STORY: MULTIPLE POINTS OF VIEW

*Be(e)hind the Story* imagines a pot of honey bearing a label that can be scanned using a phone. The label is not only a symbol, but the drawing of a bee which asks directly: "Have you ever wondered how honey gets from me to you?". When the label is scanned, an AR environment appears on the screen, inviting the user to select one of the points of view involved in the production and distribution of the honey.

The unique stories included involve humans, such as the beekeepers, but also other animal species such as bees. In addition, the user will be able to focus on the honey's point of view itself, including its distribution process. These are presented as videos sprinkled with moments of choice where users can decide whether they want more information on different issues, from animal welfare to carbon emissions. Overall, the system departs from statistical information to weave unique stories, as well as an overarching one, which are then tailored to the consumer according to their own preferences and other circumstances (e.g., their country, representing an end point of the distribution process that may be very different from others).

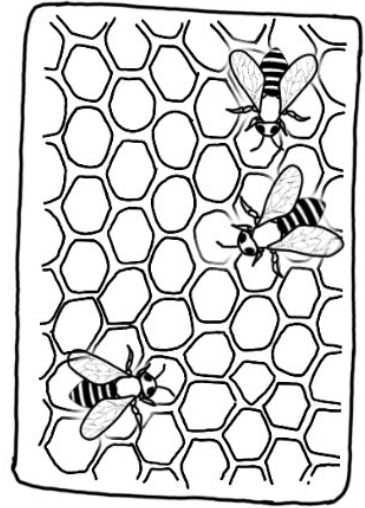

This tool is based on, and promotes, a radically transparent way of sharing sustainability-relevant indicators of animal product manufacturing, including its environmental, animal and human impact. It is not only that data is `storified', but also the actors that produce the data, and to which it refers, become `sustainability companions' for the consumer. The tool is potentially applicable to other products such as meat, whose production systems have many differences. Thus, the principle of radical transparency would be put to the test with products whose stories are 'messier' than honey's.

In this way, *Be(e)hind the Story* prioritises the consumer's autonomy, as well as their relatedness to various multi-species actors to whom they have, indeed, a deep connection: they are about to eat the fruit of their labour. By being able to choose the various stories of how the honey got to the consumer's particular city or shop, *Be(e)hind the Story* aims to provoke reflection and a deeper engagement with what is behind the scenes.

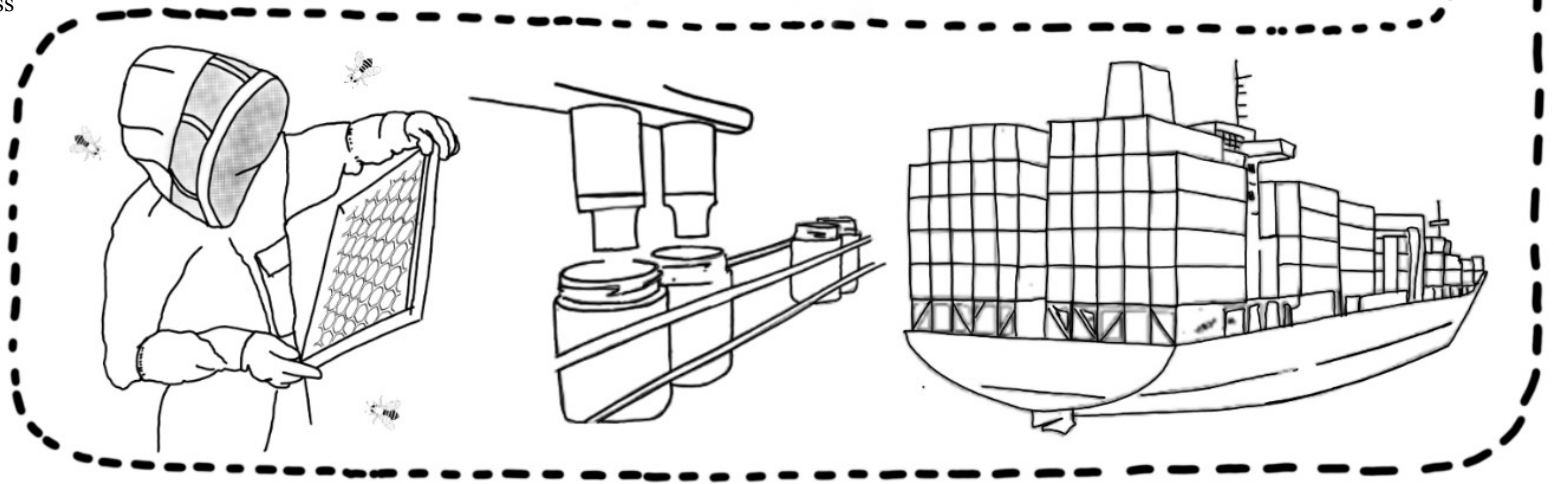

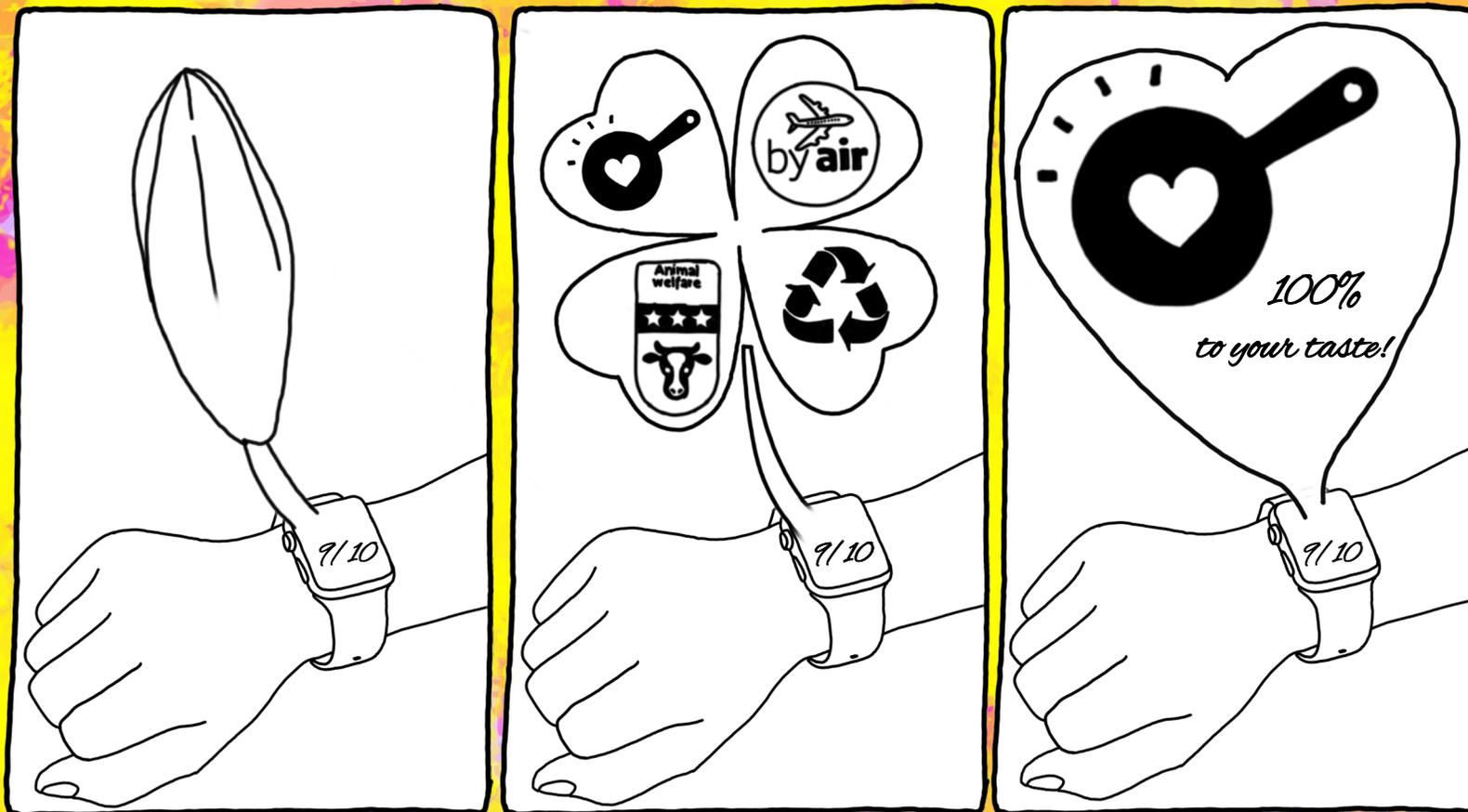

**ECO BLOOM: PERSONALISED ADAPTATION**

*Eco Bloom* provides a layered overview about how a product's qualities align with a consumer's personal pro-environmental attitudes. Upon scanning a product in the shops, a colourful closed clover appears on the consumer's smart watch or phone. The closed clover shows an overall score (1-10). Depending on the value, the label is coloured in a traffic-light metaphor. Each folded leaf, surrounding the overall score, shows a symbol of different eco-categories that are important to the consumer, i.e. animal welfare, water consumption etc. When tapping on one leaf, it folds open, revealing a detailed score for the individual category and follow-up information. Every score the consumer sees is geared towards their personal attitudes, i.e. what they consider as important and (un-)acceptable.

As pro-environmental attitudes are highly individual, *Eco Bloom* caters to the specific needs of users, providing personalised product feedback. Thereby, *Eco Bloom* adjusts to any phase of an individual's journey towards sustainability. Moreover, a key quality of this label is that it does not overwhelm the consumer with a) information unnecessary to their interests and b) too much information at once. The overall folded label fits onto the size of a smart watch display. As it is a layered approach, an overall score is the first thing consumers perceive. If a consumer desires more detailed information, they can unfold the individual clover leafs, study the sub-scores and categories that made-up the overall score and get details on-demand. Also, attitudes can always be altered through the settings, to allow adapting the score when changing opinions towards sustainability topics and categories.

Setting up *Eco Bloom* for the first time and `unfolding' information engages consumers in actively educating themselves and exploring their own attitudes towards sustainability topics. Without providing rigid `one-size-fits all' information, this type of label encourages individual perspectives and recommendations for sustainable shopping.

**A LABEL OF OUR OWN: COMMUNITY-DRIVEN ECO-LABELLING**

*A label of our own* is a smartphone app for crowdsourcing eco-labeling among consumers. It uses an AR system to enable consumers to target any product's packaging and augment it with meta-data such as: text-based annotations, drawings to hack or otherwise modify the visual appearance of the product, ratings based on consumers' perception of sustainability… By adding content to emergent AR-based eco-labels, users contribute to a collectively-owned, multi-faceted rating of products – thereby claiming shared responsibility over the fiscalisation of those products' socio-ecological impact.

Because the system is not owned by anyone in particular (the government, the retailer, the producer, the consumer), the content of the collectively crafted eco-labels can hardly be censored in a top-down manner. Rather, these eco-labels are the result of an emergent and evolving process of negotiation between consumers. As such, this design idea tackles an important issue with eco-labels: credibility. By displacing their creation from producers and retailers to consumers, it reduces chances of greenwashing.

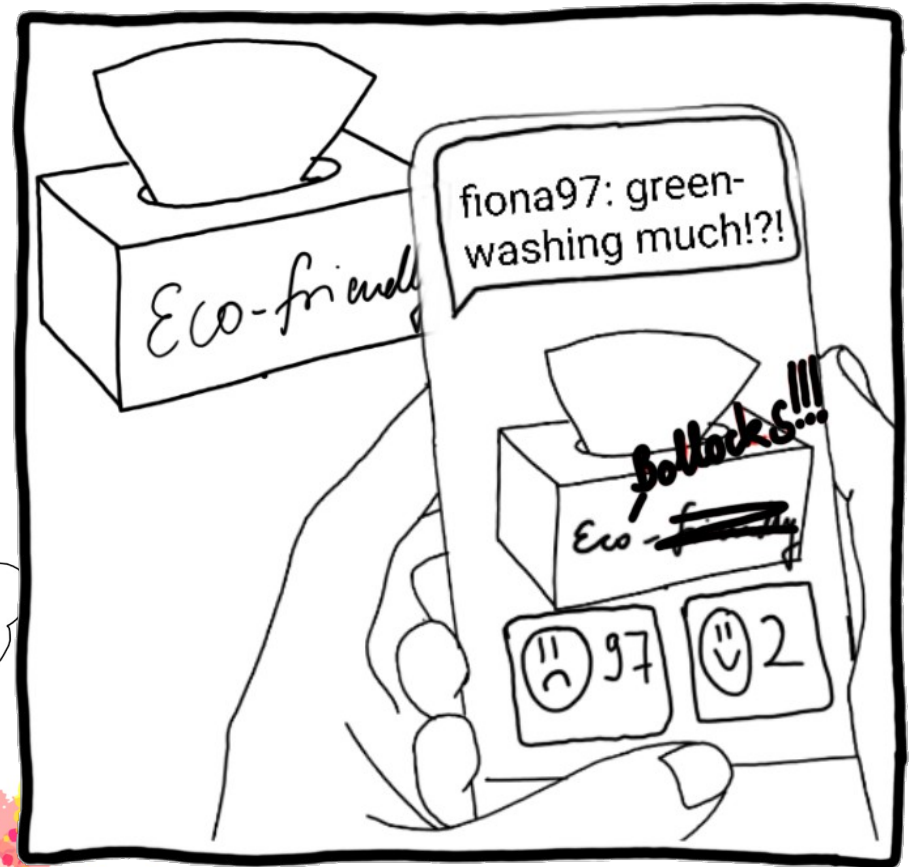

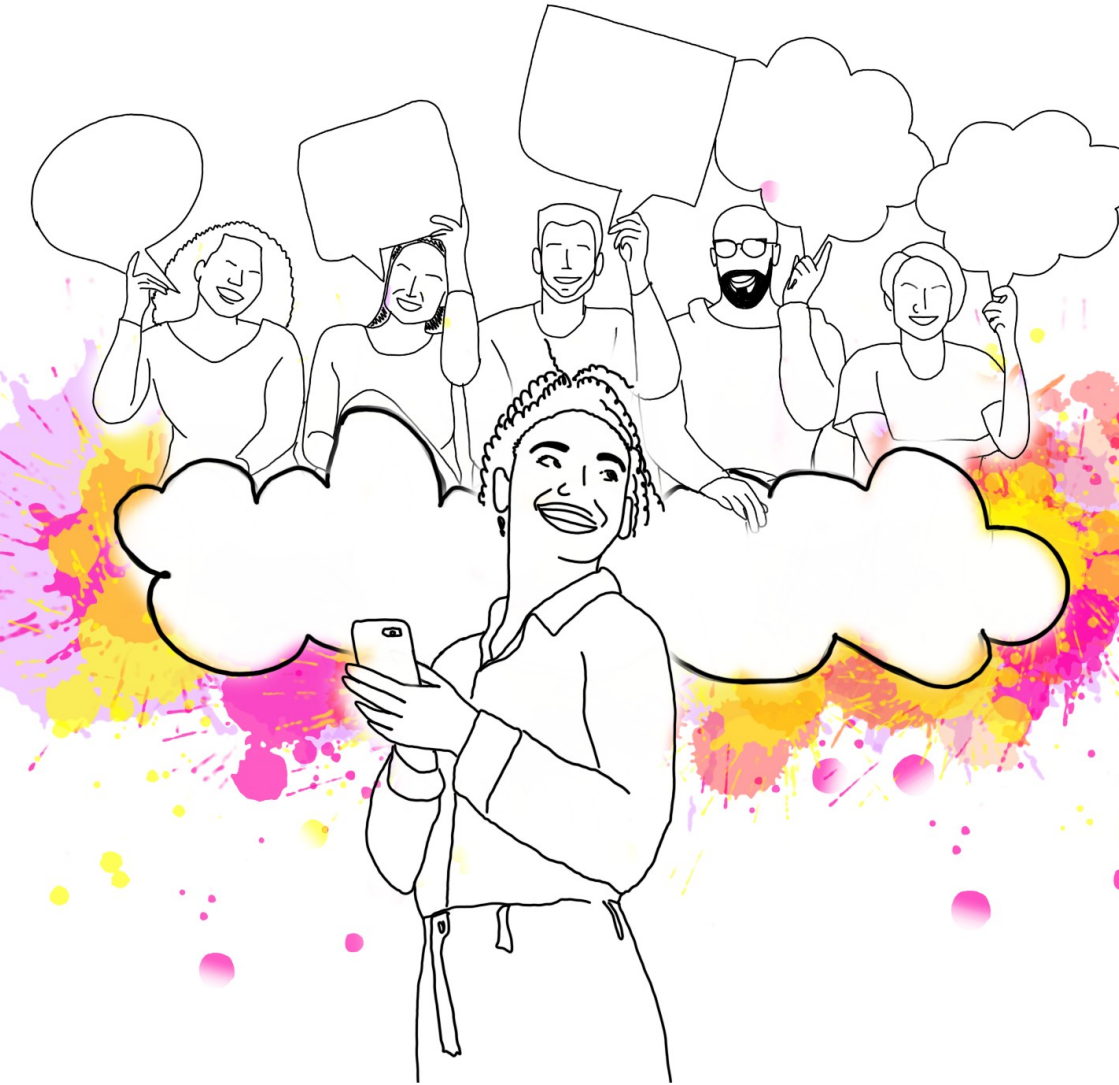

From an experiential perspective, *A label of our own* appeals to people's desire for sharing knowledge and opinion; and, conversely, from their will to learn from others'. It also taps into the joy that can derive from engaging in creative, subversive activity with an intention of making a societal impact. It builds on existing traditions of activism such as guerrilla art, which use creative practice as a form of societal transformation through subversion of the status quo. Borrowing from that principle, *A label of our own* reclaims consumers' voice when it comes to presenting products – if those products are labelled by producers in a non-truthful way, they will likely be called out by the community.

**LOUD&BOLD: ADAPTIVE PACKAGING**

*Loud&Bold* is multisensory and multimodal packaging. It is multisensory as it reacts with the content of the packaging to assess its sustainability, and it reacts with the consumer's presence and body to inform them about the sustainability of its content. It is multimodal as it provides olfactory, auditory, haptic, and visual infor-mation within different stages of consumer interaction. Consider a shopper, who wants to buy a packaged food. First, sensing the approaching shopper, *Loud&Bold* displays a "good" smell (e.g. forest) as first information. With the shopper's grasping hand inching towards the product, *Loud&Bold* plays happy and evocative music. Upon touch, the product's (etiquette's) surface colour changes from bleak to colourful, the texture changes from rough to smooth. Finally, the information on the back of the product changes, altering the written text, telling the "truth" (e.g. its origin, transportation, environmental impact), encouraging active search of additional information.

*Loud&Bold* encourages joyful discovery through layers of information. Olfactory, auditory, and haptic feedback provide high-level sustainability information. Upon active search however (i.e. turning the product around), consumers receive detailed information. The alteration of text on the back provides radical transparency increasing consumer trust, educating them, making them confident in their choice and themselves. *Loud&Bold* is a technology which invades a consumer's physical, emotional, and rational space - loudly and boldly. Simultaneously, it asks consumers to be loud and bold about their choices, be it for good or ill, as the choice is made in public.

*Loud&Bold* assesses the sustainability of its content, independent of food category and can thereby be used for any packaged foods, given the embedded technology within *Loud&Bold* is sourced sustainably, with long-term usage in mind. Through its multimodal design, *Loud&Bold* encourages consumer sustain-ability, but also raises the ques-tion of social acceptability, especially if an indivi-dual is not loud and bold, yet desires the "unsus-tainable" choice. Finally, the multimodal layers allow for disabled people to engage in everyday sustainability (which is shockingly under-explored). Providing access to participation in sustainability beyond the "perfectly sighted" population might elicit feelings of sense of control, confidence in choice, equality, and ultimately, joy among disabled communities, therefore greater sustainability, environmentally and socially.

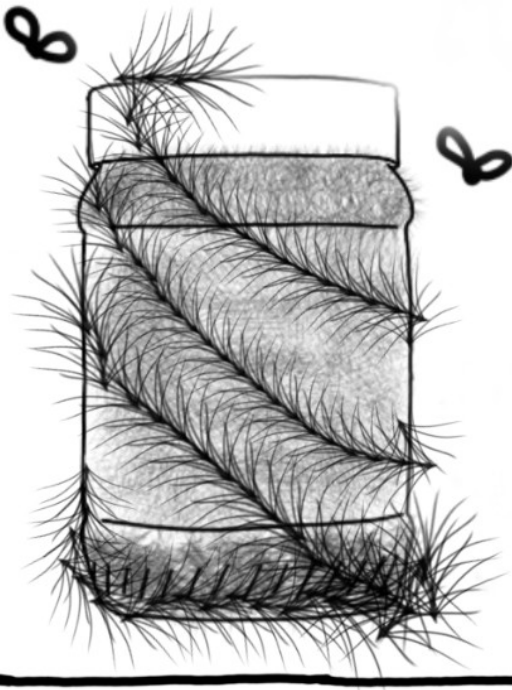
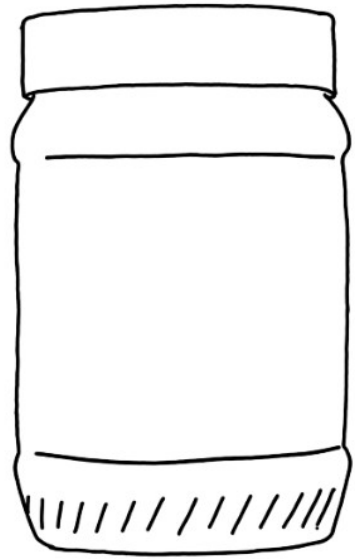
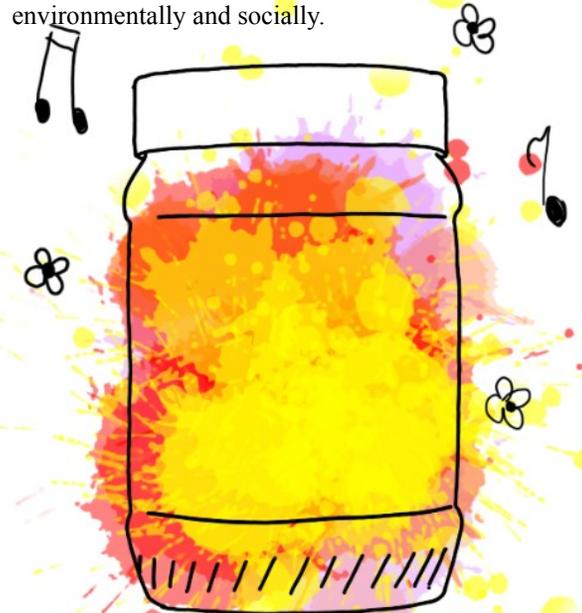

## UNPACKAGED: DISRUPTIVE PACKAGING

*Unpackaged* represents a retail system in which packaging and labels are intertwined. In it, producers are only allowed to design and use aesthetically pleasing packaging designs if their product meets all the necessary eco-certifications. The *Unpackaged* eco-labels are graphical symbols imprinted on the package, but they can also be scanned using a digital device such as a mobile phone to show, via augmented reality, the story behind the product.

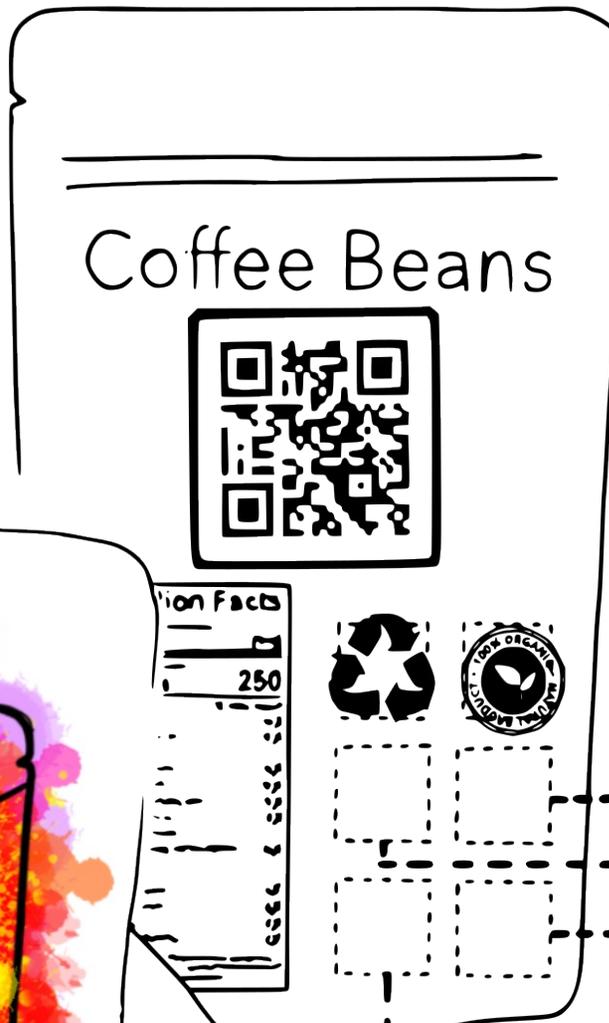

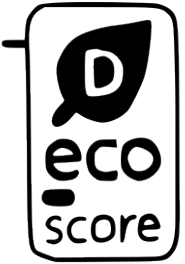

The other side of this system are the products that do not comply with the certifications required to have an attractive presentation. In this case, products must come in bland, homogeneous packaging. In addition, the absence of the necessary eco-labels is also highlighted in the packaging, showing an empty space where they should be. Since this is a status-quo disrupting system, it would likely need strong support and implementation at the policy level, similarly to tobacco and alcohol marketing.

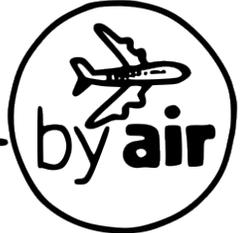

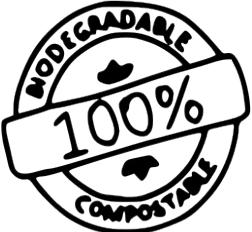

*Unpackaged* aims to provoke curiosity through its bland packaging: Why do all these products look the same? After this first impression, the consumer could choose to engage in a closer exploration, in which they would easily find that eco-labels are missing. Once again, this absence aims to stimulate further learning about the reason for this absence.

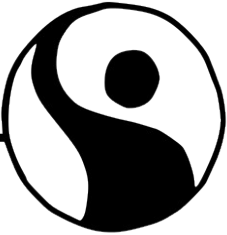

In the opposite way, the products that comply with the standards and are allowed to have attractive packaging feature eco-labels that can be explored further. The aim of using augmented reality that focuses on the product's journey aims to use an engaging medium to present a label not just as a means for conveying information, but also to create a closer relationship to the customer potentially involving self-transcendent social and environmental values. To invoke these values, the relevant aspects of the product will include not only, e.g., the materials involved in its production and the countries it has been through, but also the people who were involved. This aims to not only emphasise the physical impact of production, but also the effort required for a product to be sustainable.

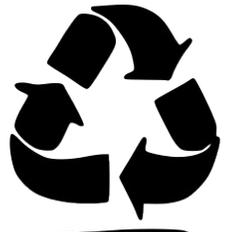

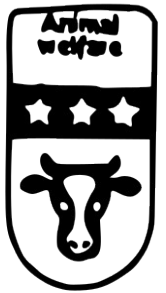

# DISCUSSION

Joy is an essential component in productive sustainability communication. This pictorial presents five anno-tated speculative ideas, surfaces three themes in which joy takes space, and derives design features from these themes which may support joyful interaction with sus-tainability. The three intertwined yet distinct themes in the presented designs include: joy in the **self**, the **other**, and **beyond** it all. The *self* describes the relationship consumers have with themselves (e.g. belief in one's own efficacy), the *other* describes the product, its re-lation to social and physical world (e.g. its story, im-pact), and *beyond* describes the interaction beyond the individual, the product, towards a holistic, long-term, and societal understanding of joy. These three themes feed into each other in an iterative fashion, potentially reinforcing socially and environmentally sustainable action across time. Although this distinction is artificial, we believe it both reflects the current state of research in this area as well as provides a useful framework within which to better understand how joy, digital eco-labels, and sustainability engagement intersect. It is important to remember that neither of the themes nor the indi-vidual design features are more important than any other. Joy is a situational, cultural, and individual con-cept and must therefore be expressed in multifaceted ways.

## Joy in the Self

One of the common themes we identified within the speculative eco-label designs is reclaiming control by diminishing feelings of powerlessness in the face of contemporary ecological crises. For instance, *Loud&Bold*, *A label of our own*, *Be(e)hind the Story*, and *Eco Bloom* empower consumers by giving them control over the extent of interaction with the tool, the type and availability of information, or the expression of data. Further, *Eco Bloom* and *Be(e)hind the Story* adapt to individuals' journey, catering customisable informa-tion, not dismissing different wants and needs but embracing them. This may increase perceived self-efficacy, and a sense of pride, and hope about the future. Consequently, and in line with literature, we identified following features which may be important to spark joy in the self.

*Customisable*. The availability of diverse data gives people the freedom to make choices on environmental as well as social issues of the food system [26], supporting the personalisation of what are the 'right' values for consumers, rather than enforcing a normative understanding. Adaptive tools allow for long-term engagement with sustainability.

*Accessible*. Accessible labels empower marginalised groups to participate in sustainable consumption [64], fostering feelings of self-efficacy and pride.

*Educational*. Educating consumers on the environ-mental impact of their choices may be key to bridging the intention-behaviour gap [60], and ultimately in tackling climate change [33].

## Joy in the Other

Label credibility and transparency are important factors that impact the joy consumers perceive. Accordingly, each proposed design emphasises the importance of radical transparency and consequently trust in the product, its story, and its agenda. The availability of data is thereby imperative, its actual use is optional. This aspect of absolute data availability refers back to the consumer's sense of autonomy, as individuals can decide whether to interact with the data or not. Access to sustainability information affords the freedom to explore, without imposing it. Further, *Be(e)hind the Story*, *Loud&Bold*, *A label of our own*, and *Unpackaged* depart from traditional interaction with groceries by either physically reacting to the individual's body, inviting the consumer on a narrative journey with the product, or going as far as to allow the rewriting of the food's story. Thereby, giving the consumer the choice to interact with data in various ways. To spark joy in the product, the following factors may be important.

*Transparent*. Label transparency fosters consumer trust and a (shared) feeling of responsibility, positively impacting and sustaining pro-environmental behaviour compliance [22].

*Tiered*. Providing tiered information encourages active search for a product's details [24]. This in turn increases consumer confidence in their knowledge and decision making and has a strong impact on consumers' willingness to buy low impact items [6, 18].

*Joyful*. Joyful interaction may increase consumer attention and responsiveness, and provide a powerful source of motivation to take action [27, 35, 45, 61]. For example, stories and narrative transportation engage climate change action through emotional arousal, inciting pro-environmental behaviour [35].

## Joy in the Beyond

Finally, the self and the other are situated in a larger societal, environmental, and economic context. For holistic, long-term views on sustainable food shopping, joy must extend beyond metrics, beyond the individual, beyond quotidian action in an attempt to shift narratives towards global empathy and action. *Be(e)hind the Story*, *Unpackaged*, and *A label of our own* go beyond providing quantitative information and attempt to instil greater empathy towards other (non-human) actors, eliciting joy through feelings of belonging, pride over accomplishment, and collective action. *Loud&Bold* extends the circle of actors empowering visually im-paired consumers in participation, for greater environ-mental and social sustainability.

*Community-driven*. Embedding community knowledge into sustainability support systems increases trust in the products and the systems, as 'normal consumers' are believed to provide the "truth" [23] and have less exploitative intentions compared to corporations. Further, community-driven activism may instil a sense of social identity [30], increase perceived self-efficacy, and foster global empathy.

**Implications and Limitations**

When deciding to develop technology-mediated solutions, the environmental footprint required must be considered in addition to the purpose of the technology. Reusable, repairable, and recyclable components should be prioritised (e.g. *Loud&Bold*, *Unpackaged*). Designs that utilise already existing hardware (e.g. *Eco Bloom*, *A label of our own*, *Be(e)hind the Story*) may have an advantage regarding their overall footprint, however, low energy consumption must be considered during all stages of development.

The consideration of who should propose, develop, or even enforce the ideas proposed leads to various different paths. *A label of our own*, for example, seems inextricably tied to bottom-up action, where systemic disruption and mostly unregulated and free participation is at the core of the idea. Similarly, *Loud&Bold* and *Unpackaged* could be implemented as a series of guerrilla actions, given their adaptation to and display of positive and negative product features. Seeking producer consent and active participation may be seen as undermining the effort itself. Alternatively, these systems could be adopted via regulated mandates resulting from overwhelming public demand. In contrast, *Be(e)hind the Story* suggests a system where producers themselves willingly open up their processes for consumers to see. *Eco Bloom*, on the other hand, could be maintained either by industry actors or by an external agent, either a company, an association or a public entity, responsible for the app and curating a database of products.

Finally, *Be(e)hind the Story*, *A label of our own*, and *Unpackaged* augment reality itself, where technology's role is to intervene and augment the user experience and support personal needs that are both utilitarian and hedonistic. The speculative ideas in this work attempt to find balance between the functional and experiential qualities, reducing the "violent" tension between caring for the self and the environment [43]. Changing the narrative is essential in creating global empathy and a societal shift; preferably away from individual consumers to systemic drivers.

**Future Work**

This work presents first speculative ideas on sustainability support when grocery shopping, with joy as the focus. We acknowledge that there are many more approaches to evoking joy, and hope that the HCI community will use this contribution as a resource to spark discussions around this complex topic. We further acknowledge that the participants' and authors' cultural perspectives may have informed what we deemed joyful and thereby guided a (limited) interpretation of joyful designs. We intend to explore and expand our understanding of joyful technology design by embracing different cultural perspectives. To create a societal shift and global empathy, we must diversify the set of joyful ideas by organising more workshops with different stakeholders, giving us diverse data to analyse and explore joyful sustainability support, beyond a close reading. We hope to have planted the seed of joy in this portfolio so that we can collectively explore how joy can guide technology design.

**CONCLUSIONS**

Positive emotions play an important role in shaping people's engagement with sustainable food. As a first step in this direction, this work contributes five speculative digital eco-labels that aim to spark joy. We discuss themes of joy in these designs and their emergent salient design features, and how these features may spark joy. The proposed ideas highlight important issues central to sustainability communication. We hope to inspire designers to 'reposition joy' as a fundamental value in engagement and to reclaim joy as an important aspect of designs promoting sustainability.

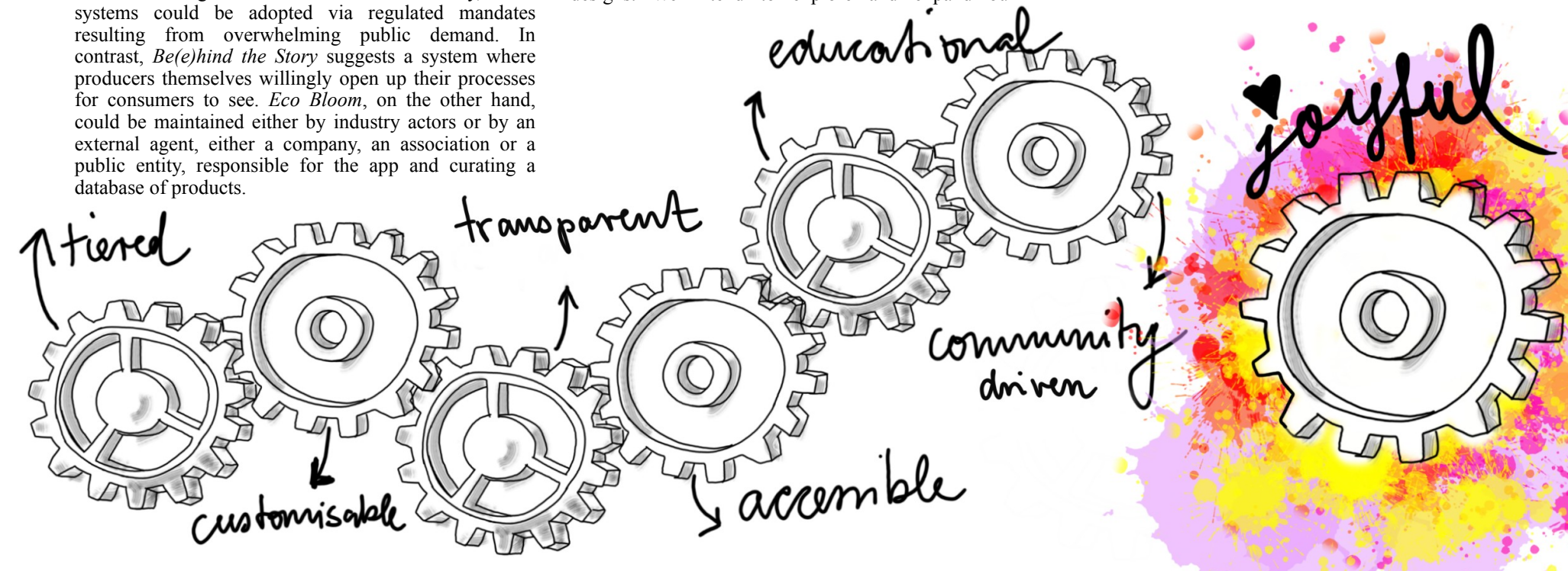